\documentclass[aps,prl,twocolumn,showpacs,superscriptaddress,floatfix]{revtex4-1}  
\usepackage[english]{babel}
\usepackage{graphicx}
\usepackage{dcolumn}
\usepackage{amsmath}
\usepackage{natbib}
\usepackage{bm}

\begin{document}

\title{Experimental approval of the extended flat bands and gapped subbands in rhombohedral multilayer graphene}
\author{Y. Henni}
\affiliation{Laboratoire National des Champs Magnétiques Intenses, CNRS, (UJF, UPS, INSA), BP 166, 38042 Grenoble, Cedex 9, France}
\author{H. P. Ojeda Collado}
\affiliation{Centro Atómico Bariloche and Instituto Balseiro, Comisión Nacional de Energía Atómica, 8400 S. C. de Bariloche, Argentina}
\affiliation{Consejo Nacional de Investigaciones Científicas y Técnicas (CONICET), Argentina}
\author{K. Nogajewski}
\affiliation{Laboratoire National des Champs Magnétiques Intenses, CNRS, (UJF, UPS, INSA), BP 166, 38042 Grenoble, Cedex 9, France}
\author{M. R. Molas}
\affiliation{Laboratoire National des Champs Magnétiques Intenses, CNRS, (UJF, UPS, INSA), BP 166, 38042 Grenoble, Cedex 9, France}
\author{G. Usaj}
\affiliation{Centro Atómico Bariloche and Instituto Balseiro, Comisión Nacional de Energía Atómica, 8400 S. C. de Bariloche, Argentina}
\affiliation{Consejo Nacional de Investigaciones Científicas y Técnicas (CONICET), Argentina}
\author{C. A. Balseiro}
\affiliation{Centro Atómico Bariloche and Instituto Balseiro, Comisión Nacional de Energía Atómica, 8400 S. C. de Bariloche, Argentina}
\affiliation{Consejo Nacional de Investigaciones Científicas y Técnicas (CONICET), Argentina}
\author{M. Orlita}
\affiliation{Laboratoire National des Champs Magnétiques Intenses, CNRS, (UJF, UPS, INSA), BP 166, 38042 Grenoble, Cedex 9, France}
\author{M. Potemski}
\email{marek.potemski@lncmi.cnrs.fr}
\affiliation{Laboratoire National des Champs Magnétiques Intenses, CNRS, (UJF, UPS, INSA), BP 166, 38042 Grenoble, Cedex 9, France}
\date{\today}
\author{C. Faugeras}
\email{clement.faugeras@lncmi.cnrs.fr}
\affiliation{Laboratoire National des Champs Magnétiques Intenses, CNRS, (UJF, UPS, INSA), BP 166, 38042 Grenoble, Cedex 9, France}

\begin{abstract}
Graphene layers are known to stack in two stable configurations,
namely ABA or ABC stacking, with drastically distinct electronic
properties. Unlike the ABA stacking, little has been done to
experimentally investigate the electronic properties of ABC
graphene multilayers. Here, we report the first magneto optical
study of a large ABC domain in a graphene multilayers flake, with
ABC sequences exceeding 17 graphene sheets.  The ABC-stacked
multilayers can be fingerprinted with a characteristic electronic
Raman scattering response, which persists even at room
temperatures. Tracing the magnetic field evolution of the inter
Landau level excitations from this domain gives strong evidence to
the existence of a dispersionless electronic band near the Fermi
level, characteristic of such stacking. Our findings present a
simple yet powerful approach to probe ABC stacking in graphene
multilayer flakes, where this highly degenerated band appears as
an appealing candidate to host strongly correlated states.
\end{abstract}

\maketitle

Tailoring the electronic and optical properties of layered
materials by controlling the layer orientation or their stacking
order is an important possibility offered by the physics of two
dimensional systems. Graphene is the first isolated two
dimensional crystal~\cite{Novoselov2004} and the properties of
graphene stacks, from mono- to multi-layers (N-LG), have been
intensively investigated in the last decade~\cite{Neto2009}. The
thermodynamically stable stacking of multilayer graphene is the
Bernal stacking, where the A sublattice in one layer comes right
below the B sublattice in the other layer~\cite{Koshino2013}. It
was not until recently that experiments on rhombohedral stacking,
with an ABC layer sequence, have been reported. ABC tri-layer
graphene, the simplest rhombohedral N-LG, have been successfully
isolated and presents a tunable band gap~\cite{Lui2011} as well as
chiral quasi-particles as evidenced from their unconventional
quantum Hall effect~\cite{Bao2011,Zhang2011,Kumar2011}.

To our knowledge, there are yet no investigations of ABC graphene
multilayers. This is mainly because of the low abundance of this
polytype in real samples. Hence, tracing the evolution of the
electronic properties of ABC-stacked multilayers when increasing
the number of layers remains challenging. One of the intriguing
predictions about electronic properties of ABC-stacked multilayers
is that they are expected to host surface states (localized mainly
on the top and bottom layers) with a flat dispersion near the
corners of the Brillouin zone, and bulk states with a band gap.
When increasing the number of ABC-stacked layers, the extend of
the flat dispersion of the low energy bands increases, while the
size of the bulk band gap decreases~\cite{Burkov2011,Xiao2011}.

\begin{figure*}
\begin{center}
\includegraphics[width=1\textwidth]{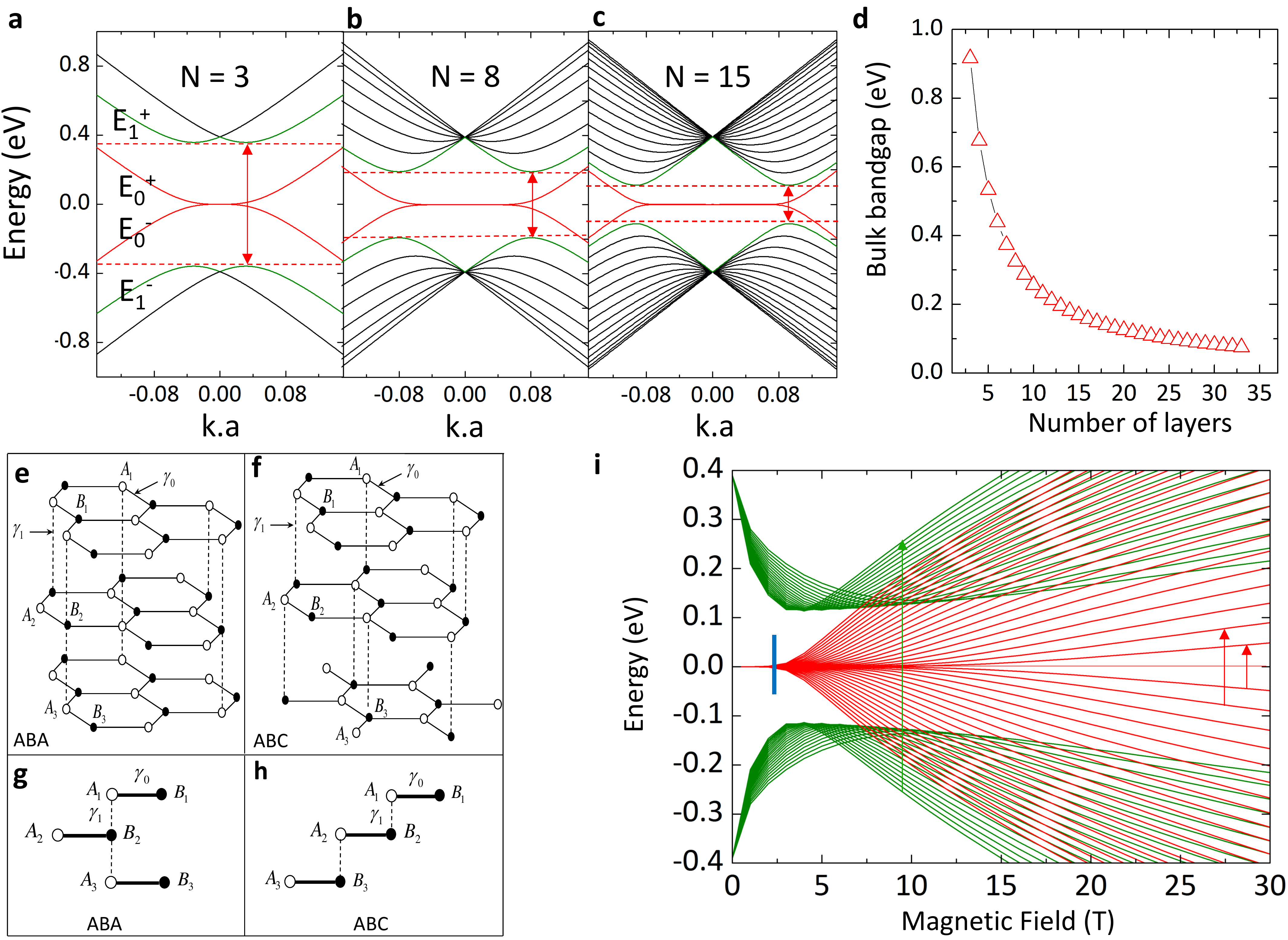}
\caption{a-c) Electronic dispersions for $N=3,8,15$ layers
obtained from the low energy effective Hamiltonian, respectively.
The dashed horizontal red lines and the arrows indicate the energy
gap. d) Evolution of the energy gap as a function of the number of
ABC-stacked layers. e,f) Schematics of the crystal structure of
ABA and ABC N-LG. Open circles and black dots are carbon atoms of
the A and B sublattices, respectively. g,h) Side views of the unit
cells of ABA and of ABC N-LG. i) Calculated dispersion of the 20
first Landau levels for the flat band (red) and for the lowest
energy bands in the bulk (green), as a function of B, for N=15
ABC-stacked layers. } \label{fig:bs}
\end{center}
\end{figure*}

Notably, the surface states of ABC-stacked graphene multilayers
have been reported as topologically protected, due to the symmetry
of this material~\cite{Xiao2011}, and may thus resemble those
well-known surface states of 3D topological
insulators~\cite{Hsieh2008,Zhang2009,Hasan2010,Qi2011}.
Interestingly, due to weak spin-orbit interaction in carbon-based
systems, an even closer analogy is found with the surface states
of topological crystalline insulators~\cite{Fu2011}. Nevertheless,
in contrast to these two families of topological insulators, the
bulk band gap of ABC-stacked graphene multilayers gradually closes
with the increasing number of atomic sheets and bulk rhombohedral
graphite may behave as a 3D topologically protected
semimetal~\cite{Ho2016}. The expected high degeneracy of the flat
bands could lead to exotic electronic ground states, such as
magnetically ordered phases or surface
superconductivity~\cite{Munoz2013,Kopnin2013,Olsen2013}.
Experimental signatures of this flat bands has been recently
observed in scanning tunnelling spectroscopy (STS) and
angular-resolved photo emission spectroscopy (ARPES)
investigations of nanometer scale domains of ABC-stacked graphene
multilayers (5 layers) grown on 3C-SiC~\cite{Pierucci2015}.

In this Letter, we demonstrate, using Raman scattering techniques,
that ABC-stacked multilayers ($N>5$) can be found within
exfoliated N-LG flakes. They have characteristic signatures in
their Raman scattering response, which allow for their
identification at room temperature. These signatures include a
change of the 2D band feature line shape~\cite{Chiu2011}, together
with an additional Raman scattering feature which we attribute to
electronic Raman scattering (ERS) across the band gap in the bulk.
ABC-stacked domains can thus be identified by spatially mapping
the Raman scattering response of different flakes. Experiments
performed with an applied magnetic field ($B$) reveal two distinct
series of electronic inter Landau level excitations, involving the
low energy surface flat bands, and the bulk gapped states,
respectively. Experimental results are explained in the frame of a
tight binding (TB) model of the electronic band structure of
rhombohedral N-layer thin graphite layers~\cite{Min2008,Yuan2011}.

\begin{figure*}
\begin{center}
\includegraphics[width=1\textwidth]{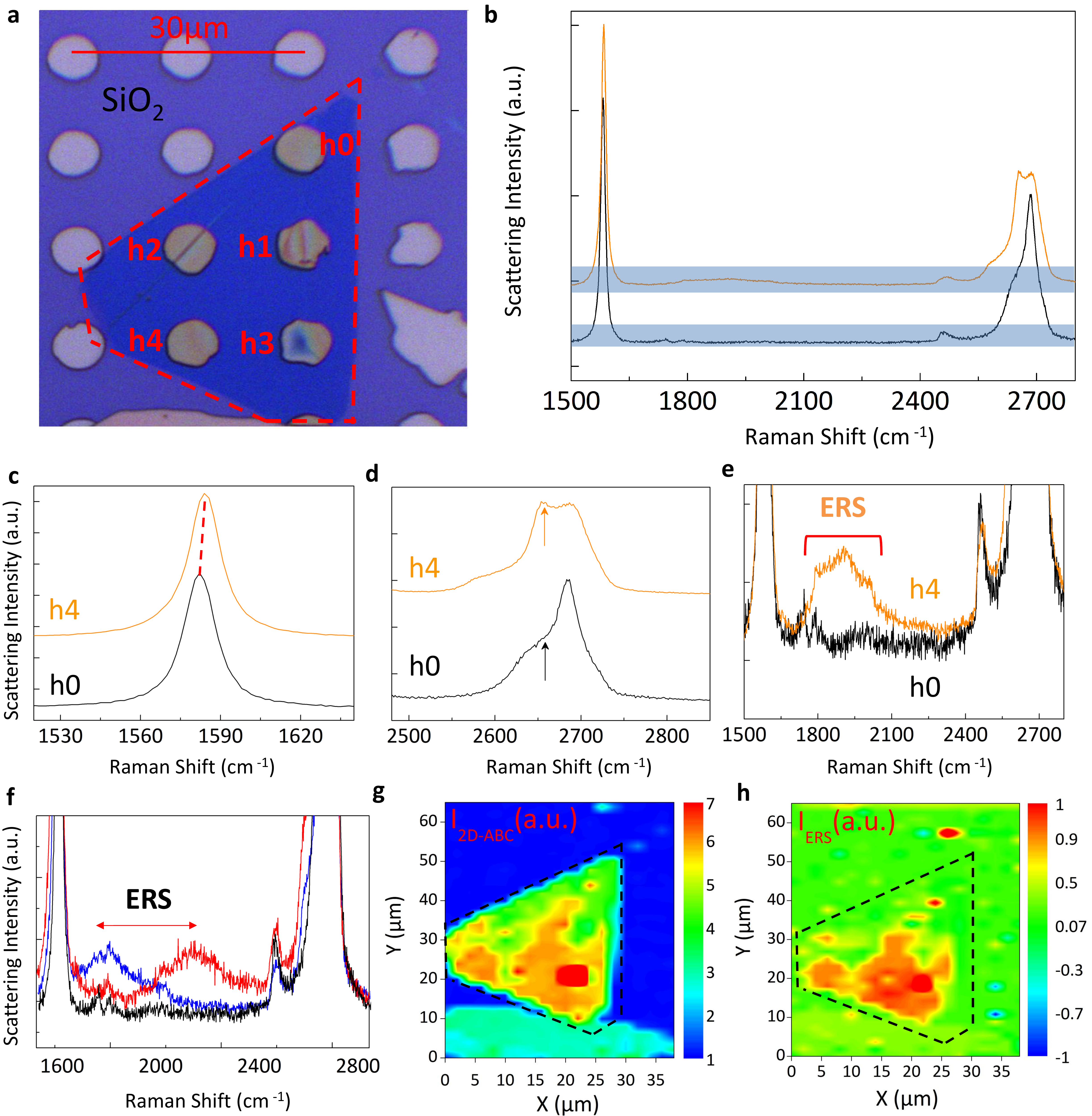}
\caption{ a) Optical microscope image of the measured flake
deposited over a silicon substrate covered with $6\mu$m regularly
spaced circular holes. The freestanding parts are labelled h0 to
h4. b) Room temperature Raman scattering spectra from h0 (black
line) and h4 (orange line). c,d,e) Zoom on details of b): G band,
2D band and on the two blue boxes depicting the low intensity
spectrum, respectively. f) Raman scattering spectra measured at
two different locations on the ABC-stacked domain (red and blue
lines) and on the AB-stacked domain (black line). g,h) False color
maps of the scattered intensity in the energy range within the 2D
band feature boxed in d), and in the energy range boxed in e)
corresponding to the ERS for a N=15 ABC-stacked layers,
respectively.}\label{fig:zero_B}
\end{center}
\end{figure*}

The 7 parameters TB model developed by Slonczewski-Weiss and
McClure~\cite{McClure1957, Slonczewski1958} for bulk graphite can
be simplified by only considering the two first intra- and inter-
nearest neighbors hopping parameters $\gamma_0$ and $\gamma_1$,
respectively (Figure~\ref{fig:bs}f-i). The calculated low energy
band structure of ABC N-LG is shown in Figure~\ref{fig:bs}a-c, for
$N = 3, 8$ and $15$ layers. Two characteristic evolutions when
increasing the number of ABC-stacked layers arise from these
calculations: (\textit{i}) the flat part of the low energy
$E_0^{\pm}$ bands (red bands in Figure~\ref{fig:bs}a-c) extends
over a larger $k$-space region, and (\textit{ii}) the energy
separation $E_g$ between the $E_1^{\pm}$ (green bands in Figure
\ref{fig:bs}a-c), decreases and ultimately closes for rhombohedral
graphite (see Figure~\ref{fig:bs}d).

A natural way of exploring electronic band structures of solids is
to apply a magnetic field in order to induce Landau quantization,
and to trace the evolution of inter Landau level excitations with
a magneto-spectroscopy techniques, such as magneto-Raman
scattering spectroscopy. The evolution with magnetic field of the
four lowest in energy bands of a $N = 15$ ABC stacked sequence is
presented in Figure~\ref{fig:bs}e. One can note that Landau levels
are formed from the flat bands (red lines), and that their
energies grow nearly linearly with the magnetic field, but
starting from a finite onset magnetic field. For $N = 15$, the
onset field is close to $B\sim 3$~T, as indicated by blue bar in
Figure~\ref{fig:bs}e. As a consequence, inter-Landau level
excitations within the flat bands (red arrows), in the first
approximation, evolve linearly with the applied magnetic field,
however with a rather unusual extrapolated negative energy offset
(see supplementary information). We anticipate that this will be
the magneto-Raman scattering signature of electronic states with
such dispersion. Landau levels in the $E^{+}_{1}$ and $E^{-}_{1}$
bands (green lines in Figure~\ref{fig:bs}e) also show a non
trivial evolution when increasing the magnetic field. From $\pm
\gamma_1$~eV at the K-point, their energies first decrease
(increase) with the magnetic field until they reach $\pm E_g/2$,
respectively. For higher magnetic fields, their energies increase
(decrease) with a quasi linear evolution. Thus, inter Landau level
excitations involving these states (green arrows) are expected to
first decrease in energy down the gap value, and to grow in a
linear way for higher magnetic fields.

An optical photograph of the investigated flake is presented in
Figure~\ref{fig:zero_B}a. The flake is produced from the
mechanical exfoliation of bulk graphite and transferred on a
SiO$_2$(90nm)/Si substrate patterned with holes in the oxide (see
Methods). The Raman scattering signal from the suspended parts is
enhanced due to an optical interference effect in the
substrate~\cite{Yoon2009,Li2012}. The N-LG flake covers five
different holes labelled h0 to h4, where it is suspended. The
flake has been characterized by atomic force microscopy (AFM)
measurements which indicate a thickness of 15 to 17 layers (see
supplementary information). Raman scattering spectra are then
measured at room temperature with 50x objective and a
$\lambda=632.8$~nm laser excitation, or at liquid helium
temperature in a solenoid using a homemade micro-magneto-Raman
scattering (MMRS) spectroscopy set-up with a $\lambda\sim785$~nm
laser excitation (see Methods). In both experiments, piezo stages
are used to move the sample with respect to the laser spot and to
spatially map the Raman scattering response of the flake, or to
investigate specific locations.

Two characteristic Raman scattering spectra measured at h0 (black
curve) and at h4 (orange curve) are presented in
Figure~\ref{fig:zero_B}b. One can recognize on these spectra the
characteristic phonon response of sp$^2$ carbon, including the G
band feature around $\sim 1580$~cm$^{-1}$, and the 2D band feature
observed around $\sim2700$~cm$^{-1}$ when measured with $632.8$~nm
excitation. If the spectrum measured at h0 corresponds to that of
thin AB-stacked layers~\cite{Faugeras2008}, the spectrum measured
at h4 is notably different, even though the thicknesses of the
layer at these two locations are similar: i) the G band energy at
h4 is blue shifted by $2-3$~cm$^{-1}$ with respect to that
measured at h0 (Figure~\ref{fig:zero_B}c, ii) the 2D band line
shape is more complex with two additional contributions indicated
by arrows in~\ref{fig:zero_B}d, and iii) an additional broad
feature (Figure~\ref{fig:zero_B}e is observed at h4 around 1805
~cm$^{-1}$ with a full width at half maximum (FWHM) of $\sim
180$~cm$^{-1}$. These are the three main Raman scattering
signatures of ABC-stacked graphene multilayers.

\begin{figure*}
\begin{center}
\includegraphics[width=1\textwidth]{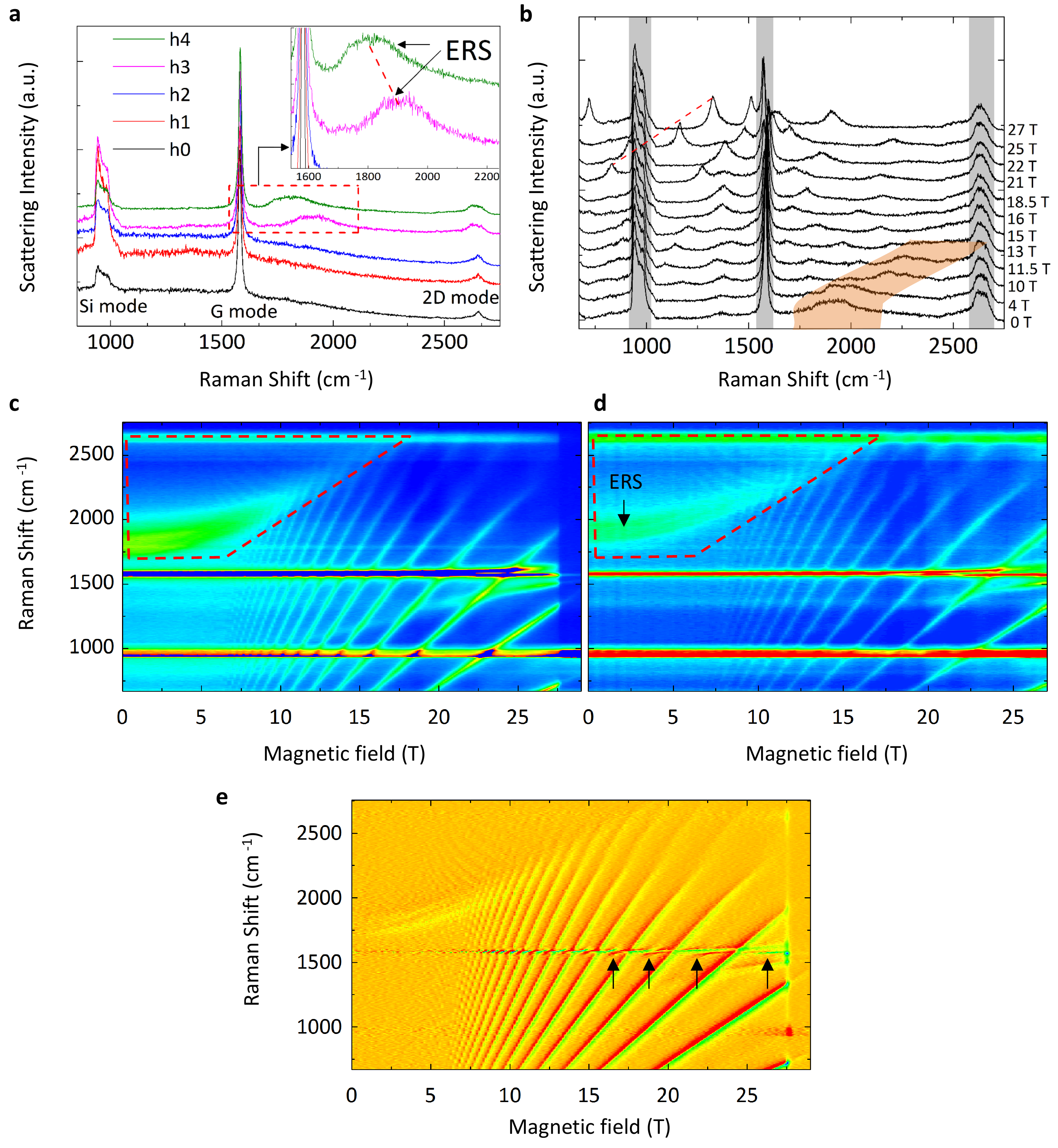}
\caption{ a) Low temperature Raman scattering spectra from the
freestanding parts on the flake. The inset is a zoom on the boxed
region. b) Raw Raman scattering spectra measured at h4 for
different values of B, showing electronic excitations from the
flat bands (red dashed line is guide for the eye) and the
evolution of the ERS (the colored zone is a guide for the eye).
The grey vertical bars in b) indicate residual contributions from
the silicon substrate, from the G band and from the 2D band
features. c,d) False color maps of the scattered intensity as a
function of the magnetic field at h3 and h4, respectively. e)
False color map of the B-differentiated data presented in d). The
arrows indicate splittings of the G band feature resulting from
the magneto-phonon resonance.} \label{fig:B_spec}
\end{center}
\end{figure*}

To study the homogeneity of the N-LG flake, we have performed
spatial mappings of the Raman scattering response at room
temperature, with $2\mu$m spatial steps. When scanning the surface
of the flake, it appears that the broad Raman scattering feature
is observed in a large domain and that its central position can
change from one location to another (see
Figure~\ref{fig:zero_B}f). As will be clarified in the following
by the magneto-Raman measurements, we interpret this feature as
arising from an electronic excitation between two $E_1^{\pm}$
bands of the band structure of ABC N-LG. Similar to metallic
carbon nanotubes~\cite{Farhat2011} and bulk
graphite~\cite{Ponosov2015}, electronic excitations do contribute
to the room temperature Raman scattering response of ABC-stacked
N-LG. The different locations on the flake where this ERS signal
is observed are presented in the form of a false color spatial map
in Figure~\ref{fig:zero_B}h. The modified line shape of the 2D
band is also observed at different locations on our flake, shown
in Figure~\ref{fig:zero_B}g. The correlation between these two
false color maps is a strong indication that the observation of
the ERS feature and of the modified 2D band line shape are
signatures of the same, ABC stacking. This flake is hence composed
of two distinct domains: the first is ABA-stacked, and is
extending over h0, h1 and h2, while the other is ABC-stacked,
extending over h3 and h4. The energy of this ERS (i.e., the bulk
band gap in Figure~\ref{fig:bs}a-c depends on the number of
ABC-stacked layers, and, taking the extreme values of the ERS
energy from the measured Raman spectra all over the ABC domain, we
estimate the number of ABC-stacked graphene layers to vary from
$\sim12$ to $\sim17$ over the whole flake. This feature is also
observed at low temperature, as can be seen in
Figure~\ref{fig:B_spec}a.

\begin{figure*}
\begin{center}
\includegraphics[width=1\textwidth]{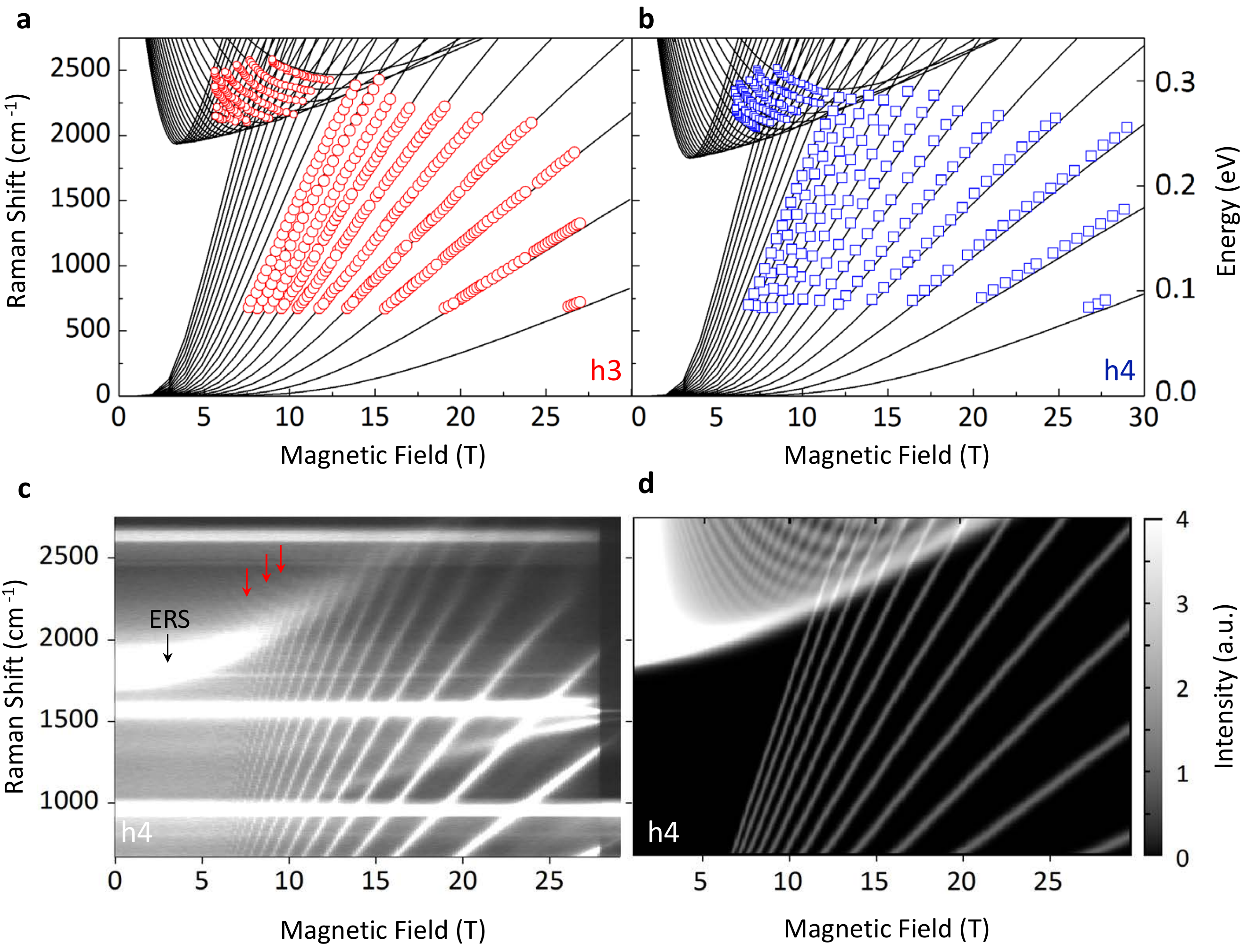}
\caption{ a-b) Evolution of the inter Landau level excitations as
a function of the magnetic field (red circles and blue squares)
together with the corresponding calculated excitation spectra for
locations h3 with N=14 and h4 with N=15, respectively.
Experimental errors are smaller than the symbol size. c) False
color map of scattered intensity at h4 as a function of the
magnetic field (the scale is saturated to clearly see the smallest
features). d) Calculated electronic excitation spectrum from both
the flat bands and the gapped bands, as a function of the magnetic
field.}\label{fig:fits}
\end{center}
\end{figure*}

Characteristic MMRS spectra measured at h4 are presented in
Figure~\ref{fig:B_spec}b, for different values of B. The central
position of the broad ERS feature increases with increasing
magnetic field and its intensity seems to vanish for $B>15$~T. The
fact that magnetic fields can influence so much the energy and
amplitude of this feature is in line with an electronic origin for
this excitation. Above $B>5$~T a series of sharp features,
dispersing with the magnetic field, appears in the Raman
scattering response. They have a symmetric line shape, in contrast
to electronic excitations observed in bulk
graphite~\cite{Kossacki2011}.

The evolution of the Raman scattering response with magnetic
field, measured at h3 and h4, are presented in
Figures~\ref{fig:B_spec}c-d, respectively, in the form of false
color maps of the scattered intensity. These two locations show a
similar response. The observed electronic excitation spectrum in
magnetic field is composed of a series of features linearly
dispersing with increasing magnetic field and of the ERS observed
at $B=0$, that acquires a fine structure at finite $B$. The
linearly dispersing features, in line with magneto-Raman
scattering selection rules in graphene~\cite{Kashuba2009} or in
graphite~\cite{Kossacki2011}, can be attributed to symmetric
($\Delta|n|=0$, where n is the Landau level index) inter Landau
level excitations within the $E_0^{\pm1}$ bands and they represent
the strongest contribution to the electronic Raman scattering
spectrum in magnetic field. Optical-like excitations
($\Delta|n|=\pm1$) are not directly seen but they effectively
couple to the G band. They give rise to the magneto-phonon
resonance and to the associated anti-crossings when they are tuned
in resonance with the G band energy~\cite{Ando2007, Kashuba2009}.
Such anti-crossings are indicated by red arrows in the
B-differentiated color map of h4, presented in
Figure~\ref{fig:B_spec}f. The detailed analysis of the
magneto-phonon resonance in ABC N-LG is beyond the scope of this
paper. Of much weaker intensity, Raman scattering features with an
energy decreasing when increasing the magnetic field are observed
(red marked regions in Figure~\ref{fig:B_spec}d-e. These features
are better seen in Figure~\ref{fig:B_spec}e, that shows the $B=0$
subtracted false color map of h3. They correspond to symmetric
($\Delta|n|=0$) inter Landau level excitations within the lowest
$E_1^{\pm1}$ bands in the bulk. The energy of such excitations
first decreases with the magnetic field as long as the Landau
levels are confined in the cone of the corresponding band
structure at $B=0$ (Figure~\ref{fig:bs}c) and then, when the inter
Landau level energy spacing reaches the value of the energy gap in
the bulk, they merge with the broad ERS feature.

The evolution of these two families of electronic excitations with
increasing magnetic field is grasped by our tight-binding
analysis. We first determined the central positions of the
different observed excitations using Lorentzian functions, and
searched for the parameters entering the model to best describe
our data. From this modelling, we can determine the number of
layers to be $N=14$ and $15$ for h3 and h4, respectively
(Figures~\ref{fig:fits}a-b), while we set $\gamma_0=3.08$~eV and
$\gamma_1=0.39$~eV, as observed in bulk graphite~\cite{Nakao1976,
Kossacki2011, Kuehne2012, Berciaud2014}. In a simple approach (see
supplementary information), we calculated the electronic
excitation spectrum~\cite{Kashuba2009,Kashuba2012,Marcin2010} and
its evolution when applying a magnetic field. These results are
compared to the experimental evolution in
Figures~\ref{fig:fits}c-d for h4. The model reproduces the main
observed features. In particular, the electronic excitation
observed at $B=0$ is reproduced and arises indeed from an inter
band electronic excitation that loses its spectral weight when the
magnetic field is increased, transforming into inter Landau level
excitations with their characteristic negative energy dispersion.

We report the observation of electronic excitations in N-LG
system, which includes a large domain of $\sim15$ ABC-stacked
layers. The analysis of its low energy electronic excitations with
magnetic field can be understood in the frame of a tight-binding
model with three parameters, the number of ABC-stacked layers and
the intra- and inter-layer nearest neighbors hopping integrals
$\gamma_0$ and $\gamma_1$. Such stacking has a unique signature in
its Raman scattering response at $B=0$, in the form of a low
energy electronic excitations across the band gap in the bulk. The
ERS response is also observed at room temperature. The central
position of the ERS is related to the number of ABC-stacked layers
which can then be deduced using simple Raman scattering
spectroscopy. Our findings underscore the rich physics hidden in
graphene multilayers with ABC stacking, namely the existence of
electronic bands with a flat dispersion (diverging density of
states) localized on the surface, and of electronic states in the
bulk with an energy gap that depends on the number of layers.
These results represent an impetus for other studies targeting the
highly correlated surface states, which may lead to emergent
exotic electronic ground states on this system, such as magnetic
order or superconductivity.


\begin{acknowledgements}

This work has been supported by the ERC Advanced Grant MOMB
(contract no. 320590) and the EC Graphene Flagship (project no.
604391). H.P.O.C., G.U. and C.A.B. acknowledge financial support
from PICTs 2013-1045 and Bicentenario 2010-1060 from ANPCyT, PIP
11220110100832 from CONICET and grant 06/C415 from SeCyT-UNC.

\end{acknowledgements}

\section{Methods}
The measured flake was obtained from the mechanical exfoliation of
natural graphite. It was then transferred non-deterministically on
top of a undoped SiO$_{2}$(90nm)/Si substrate on top of which an
array of holes ($\sim6\mu$m diameter) was etched by means of
optical lithography techniques. Room temperature Raman
characterization is performed on the flake, using a helium-neon
laser source of $\lambda= 633$~nm (i.e., $E=1.95$~eV). To probe
the Landau levels dispersion in our ABC N-LG flake, we used an
experimental set-up comprising a homemade micro-Raman probe that
operates at low temperatures and high magnetic fields. The
excitation source was provided by a solid state Titanium doped
sapphire laser. The excitation wavelength fixed at  $\lambda=
785$~nm  (i.e., $E= 1.58$~eV) is brought to the sample by a
$5\mu$m core mono-mode optical fiber. The end of the Raman probe
hosts a miniaturized optical table comprising a set of filters and
lenses in order to clean and focalize the laser spot. A high
numerical aperture lens is used to focalize the laser light on the
sample, which is mounted on X$-$Y$-$Z piezo stages, allowing us to
move the sample relative to the laser spot with sub-$ \mu $m
accuracy. The non-polarized backscattered light is then injected
into a 50 $ \mu $m  multi-mode optical fiber coupled to a
mono-chromator equipped with a liquid nitrogen cooled charge
coupled device (CCD) array. The excitation laser power was set to
$ \sim $ 1 mW and focused onto a $\sim1\mu $m diameter spot. The
resulting intensity is sufficiently low to avoid significant laser
induced heating and subsequent spectral shifts of the Raman
features. The probe is then placed in a resistive magnetic
equipped with a liquid He cryostat at 4 K. The evolution of the
Raman spectrum with magnetic field was then measured on the
freestanding parts by sweeping the values of the magnetic field.
Each spectra was recorded for a $\delta B= 0.15$~T in order to
avoid any significant broadening of the magnetic dependent
features line-widths. The details of the theoretical description
of the LLs and the magneto-Raman spectrum are given in the
supplementary information.

\end{document}